\pdfoutput=1
\documentclass[a4paper,fleqn,usenatbib]{mnras}
\usepackage[T1]{fontenc}
\usepackage{ae,aecompl}
\usepackage{graphicx}	
\usepackage{amsmath}	
\usepackage{amssymb}
\usepackage{pdflscape}
\usepackage{siunitx}
\def\lya{Ly$\alpha$}
\def\lyb{Ly$\beta$}
\def\teff{$\tau_\mathrm{eff}$}

\def\HII{\hbox{H~$\scriptstyle\rm II$}}
\def\HeII{\hbox{He~$\scriptstyle\rm II$}}
\usepackage{supertabular}
\usepackage{xcolor}
\definecolor{notecolor}{rgb}{0.8,0,0}

\usepackage{times}

\title[Opacity distribution]{Large Lyman-$\alpha$ opacity fluctuations
  and low CMB $\tau$ in models of late reionization with large islands
  of neutral hydrogen extending to $z<5.5$}

\author[Kulkarni et al.]{{Girish Kulkarni$^{1,2,6}$\thanks{Email:
      kulkarni@theory.tifr.res.in},
    Laura C.~Keating$^3$,
    Martin G.~Haehnelt$^{1,2}$, Sarah E.~I.~Bosman$^4$,}
  \newauthor{Ewald Puchwein$^{1,2,7}$,
    Jonathan Chardin$^5$ and Dominique Aubert$^5$}\\
  $^1$Institute of Astronomy,
  University of Cambridge, Madingley Road, Cambridge CB3 0HA,
  UK \\
  $^2$Kavli Institute of Cosmology,
  University of Cambridge, Madingley Road, Cambridge CB3 0HA,
  UK \\
  $^3$Canadian Institute for Theoretical Astrophysics,
  60 St.~George Street, University of Toronto, ON M5S 3H8,
  Canada\\
  $^4$Department of Physics and Astronomy,
  University College London,
  London WC1E 6BT,
  UK\\
  $^5$Observatoire Astronomique de Strasbourg,
  11 rue de l'Universite,
  67000 Strasbourg,
  France\\
  $^6$Department of Theoretical Physics,
  Tata Institute of Fundamental Research,
  Homi Bhabha Road,
  Mumbai 400005,
  India\\
  $^7$Leibniz-Institut für Astrophysik Potsdam (AIP),
  An der Sternwarte 16,
  D-14482 Potsdam,
  Germany}

\date{Accepted ---. Received ---; in original form ---}

\pubyear{2018}

\begin{document}
\label{firstpage}
\pagerange{\pageref{firstpage}--\pageref{lastpage}}
\maketitle

\begin{abstract}
  High-redshift QSO spectra show large spatial fluctuations in the
  \lya\ opacity of the intergalactic medium on surprisingly large
  scales at $z\gtrsim 5.5$.  We present a radiative transfer
  simulation of cosmic reionization driven by galaxies that reproduces
  this large scatter and the rapid evolution of the \lya\ opacity
  distribution at $5<z<6$. The simulation also reproduces the low
  Thomson scattering optical depth reported by the latest CMB
  measurement and is consistent with the observed short near-zones and
  strong red damping wings in the highest-redshift QSOs.  It also
  matches the rapid disappearance of observed \lya\ emission by
  galaxies at $z\gtrsim 6$.  Reionization is complete at $z=5.3$ in
  our model, and 50\% of the volume of the Universe is ionized at
  $z=7$.  Agreement with the \lya\ forest data in such a late
  reionization model requires a rapid evolution of the ionizing
  emissivity of galaxies that peaks at $z\sim 6.8$.  The late end of
  reionization results in a large scatter in the photoionisation rate
  and the neutral hydrogen fraction at redshifts as low as $z\lesssim
  5.5$ with large residual neutral `islands' that can produce very
  long Gunn-Peterson troughs resembling those seen in the data.
\end{abstract}

\begin{keywords}
dark ages -- reionization, first stars -- intergalactic medium --
radiative transfer -- galaxies: high-redshift -- quasars: absorption
lines
\end{keywords}

\section{Introduction}

The effective optical depth ($\tau_\mathrm{eff}$) of
\mbox{Lyman-$\alpha$} (\lya) absorption in QSO spectra at redshift
$z\gtrsim 5$ is observed to exhibit large spatial fluctuations
\citep{2006AJ....132..117F, 2007AJ....134.2435W, 2015MNRAS.447.3402B,
  2017A&A...601A..16B, 2017MNRAS.466.4568T, 2018MNRAS.tmp.1287B,
  2018ApJ...864...53E}.  \citet{2015MNRAS.447.3402B} showed that the
dispersion in $\tau_\mathrm{eff}$ at these redshifts is significantly
larger than that expected from density fluctuations alone
\citep{2006ApJ...639L..47L, 2007ApJ...670...39L, 2010MNRAS.407.1328M}.
Although this suggests that the observed \lya\ data are probing a
fluctuating UV background due to patchy reionization,
\citet{2015MNRAS.447.3402B} found that the scatter in
$\tau_\mathrm{eff}$ is also greater than that in models with a
fluctuating UV background with a spatially uniform mean free path of
ionizing photons.

\begin{figure*}
  \begin{center}
    \includegraphics[width=\textwidth]{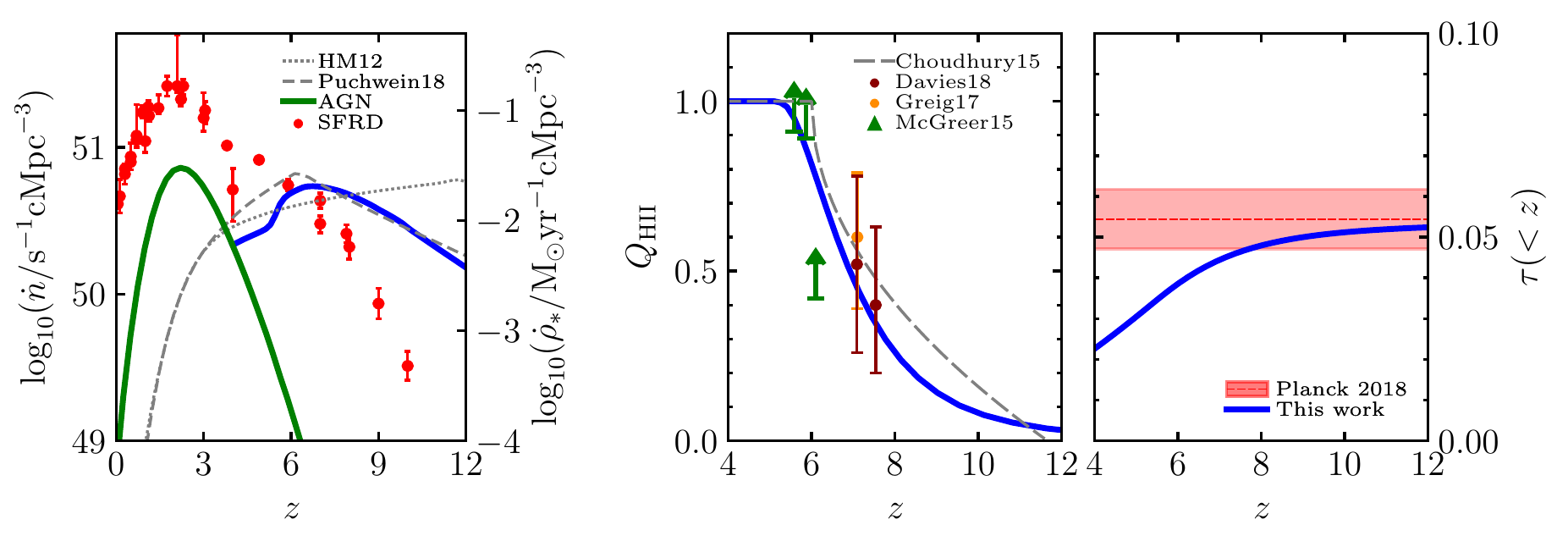}
  \end{center}
  \caption{The comoving ionizing emissivity evolution in our
    simulation is shown by the blue curve in the left panel.  Red data
    points show cosmic star formation rate density estimates at $z=9$
    and $10$ from \citet{2014ApJ...786..108O} and
    \citet{2018ApJ...855..105O} and at lower redshifts from the
    compilation by \citet{2014ARA&A..52..415M}.  The green curve shows
    the comoving ionizing emissivity from AGN brighter than
    $M_\mathrm{1450}=-21$ from \citet{2018arXiv180709774K}.  The
    ionizing emissivity due to galaxies in the models by
    \citet{2012ApJ...746..125H} and \citet{2018arXiv180104931P} are
    shown by the grey dotted and dashed curves, respectively.  The
    middle panel shows the volume-averaged ionized hydrogen fraction,
    which does not reach $Q_\mathrm{HII}=1$ before $z=5.3$.  Also
    shown are constraints on $Q_\mathrm{HII}$ by
    \citet{2017MNRAS.466.4239G} and \citet{2018arXiv180206066D} from
    the red damping wings and short near-zones in the two highest
    redshift QSOs known at $z=7.1$ and $7.5$ (with corresponding 68\%
    uncertainties), and lower limits on the ionized hydrogen fraction
    from \citet{2015MNRAS.447..499M} derived from dark pixels in
    \lya\ and \lyb\ forests.  The grey curve shows the evolution of
    $Q_\mathrm{HII}$ from the `Very Late' reionization model that
    \citet{2015MNRAS.452..261C} have shown to be consistent with the
    rapid disappearence of the \lya\ emission of high-redshift
    galaxies.  The right panel shows the electron scattering optical
    depth $\tau$, together with the \citet{2018arXiv180706209P}
    measurement of $\tau=0.0544\pm 0.0073$.}
  \label{fig:calibration}
\end{figure*}

Using a semi-numerical reionization model, \citet{2016MNRAS.460.1328D}
showed that fluctuations in the mean free path due to spatial
variation in the photoionisation rate and gas density can explain the
observed distribution of \teff\ at $z=5.6$, albeit with a rather short
mean free path that decreases rapidly with distance from (bright)
ionizing sources and that is in the mean a factor of $3.6$ smaller
than that expected from an extrapolation of measurements at
$z=4.56$--$5.16$ \citep{2014MNRAS.445.1745W}. The model by
\citet{2016MNRAS.460.1328D} also does not address the rather rapid
evolution of the \lya\ opacity distribution at $5<z<6$.
\citet{2015MNRAS.453.2943C, 2017MNRAS.465.3429C} presented a model
where the observed large \teff\ fluctuations arise due to fluctuations
in a UV background with a significant contribution from rare, bright
sources such as quasars that have a mean separation greater than the
mean free path.  Despite the resulting large fluctuations of the
photoionisation rate the ``rare-source model" of
\citet{2015MNRAS.453.2943C} struggled, however, to reproduce the long
(up to 110~cMpc$/h$) and dark ($\tau_\mathrm{eff}\gtrsim 7$)
\lya\ absorption troughs seen down to $z\sim 5.5$
\citep{2015MNRAS.447.3402B} unless the space density of
intermediate-brightness quasars is higher than that inferred from QSO
surveys \citep{2018arXiv180709774K} by a factor 3--10
\citep{2017MNRAS.465.3429C}.  The required large contribution of QSOs
to the ionizing emissivity at 1~Ry appears also to be in conflict with
the observed \HeII\ opacity and measurements of the temperature of the
IGM \citep{2017MNRAS.468.4691D, 2018arXiv180104931P} unless unlike
normal QSOs the rare ionizing sources have little emission at energies
larger than 1~Ry.  Another explanation for the large fluctuations in
\teff\ was proposed by \citet{2015ApJ...813L..38D}, who argued that
spatial variation in gas temperature can lead to the observed scatter
in \teff\ due to the temperature dependence of the recombination rate,
albeit with a reionization history that is more extended than
suggested by recent CMB and \lya\ absorption and emission data (see
\citealt{2018MNRAS.477.5501K} for a discussion).

While efforts are underway to observationally determine which of these
models describes the origin of the \teff\ fluctuations
\citep{2018ApJ...863...92B, 2018ApJ...860..155D}, it is clearly
necessary to self-consistently model spatial variation in the UV
background, the mean free path of ionizing photons, and the gas
temperature in radiative transfer simulations of reionization.  In
this Letter, we present results from such a simulation that allows us
to probe scales larger than the rapidly increasing mean free path
during the overlap of \HII\ regions.  This has become possible by
pushing our simulations to $2048^3$ particles/cells with a box size of
160~Mpc$/h$ that allow us to sample a large enough cosmological volume
at sufficient resolution while enabling us to avoid several
simplifying assumptions made in the models described above.  After
deriving the spatial distribution of the neutral hydrogen fraction,
the photoionization rate, and the gas temperature, we investigate the
distribution of \teff\ and compare it to data.  We present the details
of our simulation in Section~\ref{sec:sim}.  Section~\ref{sec:res}
discusses the \teff\ fluctuations in our model.  Our $\Lambda$CDM
cosmological model has $\Omega_\mathrm{b}=0.0482$,
$\Omega_\mathrm{m}=0.308$, $\Omega_\Lambda=0.692$, $h=0.678$,
$n_\mathrm{s}=0.961$, $\sigma_8=0.829$, and $Y_\mathrm{He}=0.24$
\citep{2014A&A...571A..16P}.

\begin{figure*}
  \begin{center}
    \includegraphics[width=\textwidth]{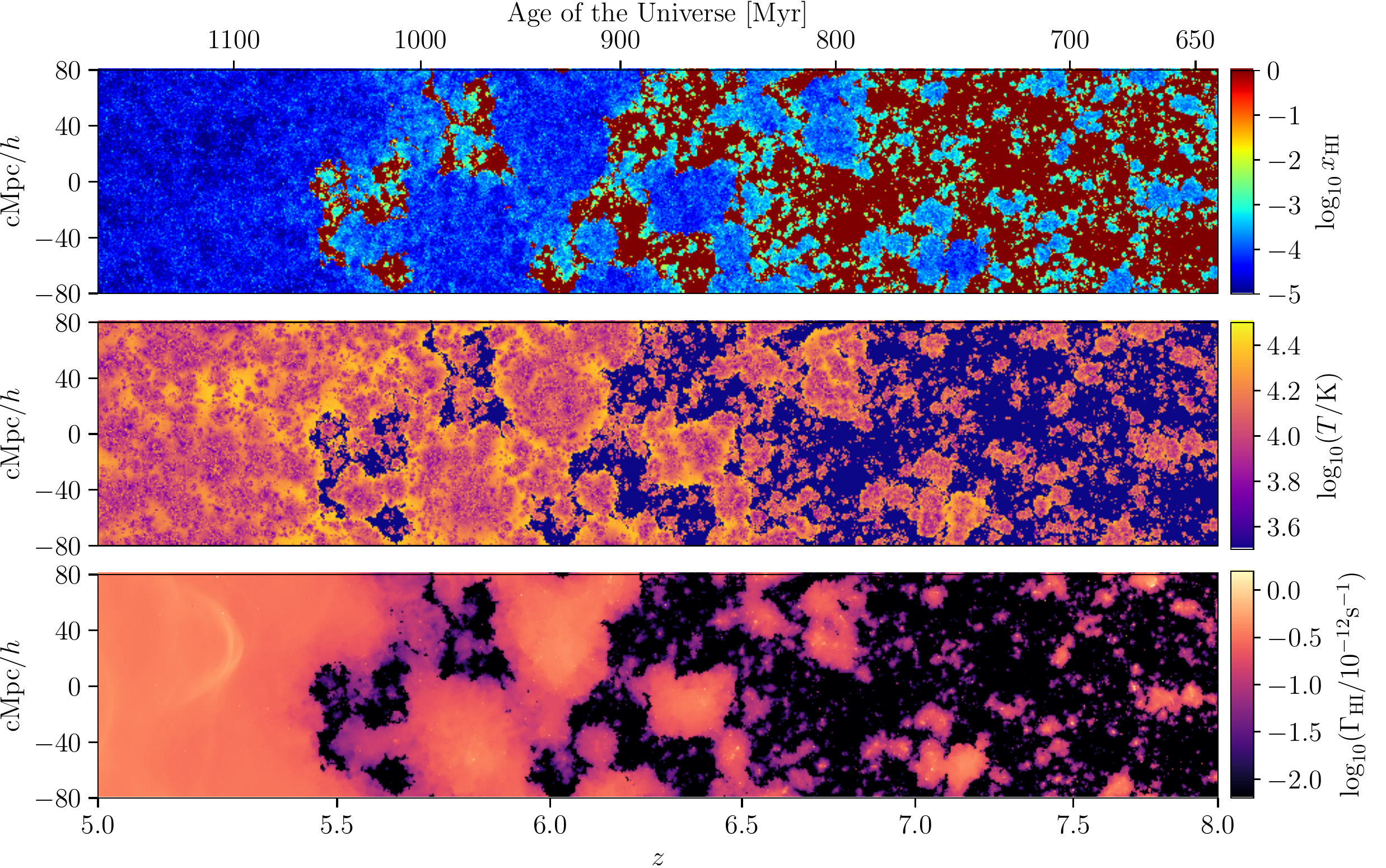}
  \end{center}
  \caption{Evolution of the neutral hydrogen fraction (top panel), gas
    temperature (middle panel), and the hydrogen photoionization rate
    (bottom panel) from $z=8$ to $5$ in the simulation.}
  \label{fig:slices}
\end{figure*}

\section{Simulation set-up and calibration}
\label{sec:sim}

We perform single-frequency cosmological radiative transfer using the
ATON code \citep{2008MNRAS.387..295A, 2010ApJ...724..244A}, following
an approach similar to \citet{2015MNRAS.453.2943C} and
\citet{2018MNRAS.477.5501K}.  Cosmological density fields obtained
from hydrodynamical simulations are post-processed by ATON.

Our cosmological hydrodynamical simulation was performed using the
\textsc{p-gadget-3} code, which is derived from the \textsc{gadget-2}
code \citep{2001NewA....6...79S, 2005MNRAS.364.1105S}.  We used a box
size of 160~cMpc$/h$ with $2048^3$ gas and dark matter particles with
a dark matter particle mass of $M_\mathrm{dm}=3.44\times 10^7$
M$_\odot/h$ and gas particle mass of $M_\mathrm{gas}=6.38\times 10^6$
M$_\odot/h$.  The initial conditions are identical to those of the
160--2048 simulation from the Sherwood simulation suite
\citep{2017MNRAS.464..897B}.  These initial conditions were evolved
from $z=99$ to $z=4$.  We saved 38 snapshots at 40~Myr intervals.  In
order to speed up the simulation, we used the {\tt QUICK\_LYALPHA}
option in \mbox{\textsc{p-gadget-3}} to convert gas particles with
temperature less than $10^5$~K and overdensity of more than a thousand
to star particles \citep{2004MNRAS.354..684V}.  This approximation
does not affect the reionization process as the mean free path of
ionizing photons is determined by self-shielded regions with a typical
overdensity of $\Delta=10$--$100$ \citep{2009MNRAS.394.1812P,
  2018MNRAS.478.1065C}.  We grid the gas density on a cartesian grid
with the number of grid cells equal to the number of SPH particles,
yielding a grid resolution of 78.125~ckpc$/h$.  ATON is then used to
perform radiative transfer in post-processing.  ATON solves the
radiative transfer equation by using a moment-based description with
the M1 approximation for the Eddington tensor
\citep{2001NewA....6..437G, 2008MNRAS.387..295A} and self-consistently
derives the fraction of ionized hydrogen and the gas temperature on
the grid.  The adiabatic cooling of gas due to cosmic expansion is
accounted for.  The hydrodynamic response of the gas due to the
changes in temperature is neglected.  However, we do not expect this
to seriously affect the results, as the pressure smoothing scale at
redshifts $z>5$ for our chosen UV background is less than 100 ckpc$/h$
\citep{2015ApJ...812...30K, 2017ApJ...837..106O}, approximately equal
to the cell size of our grid.  Ionizing sources are placed at the
centres of mass of haloes with masses above $10^9$~M$_\odot/h$.

\begin{figure*}
  \begin{center}
    \includegraphics[width=\textwidth]{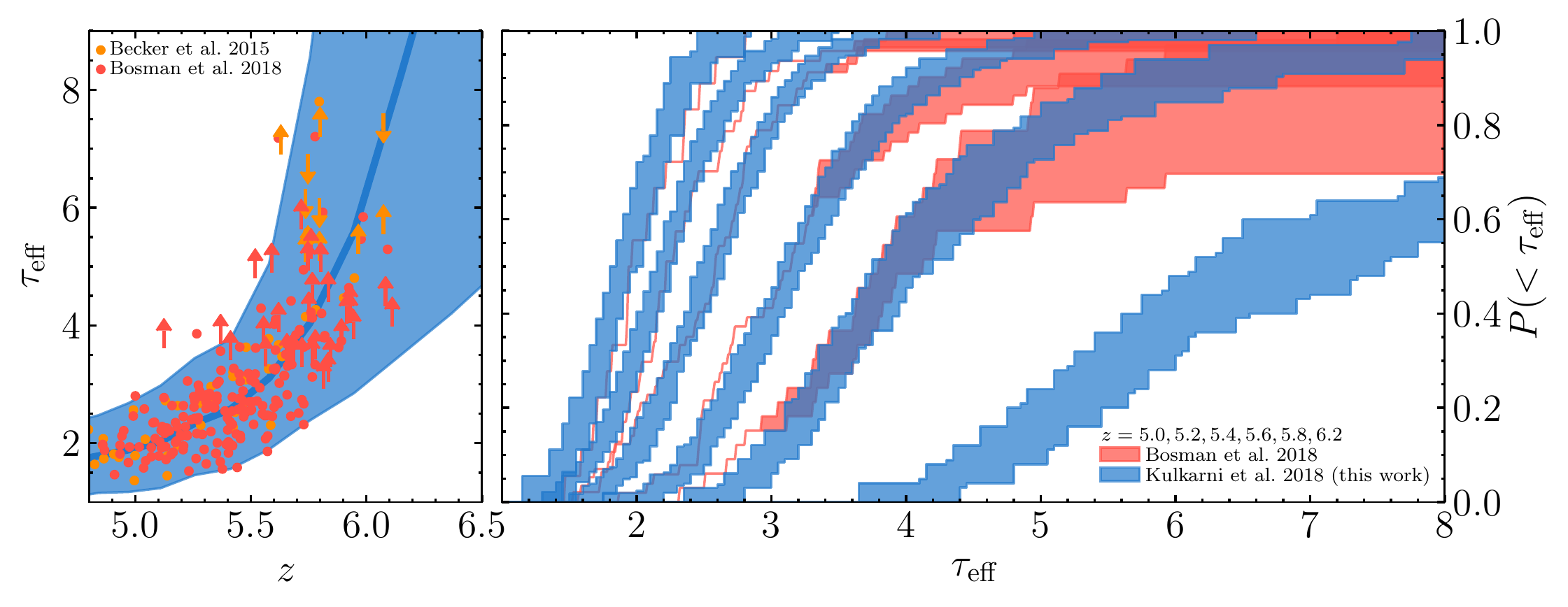}
  \end{center}
  \caption{The left panel compares the evolution of the effective
    \lya\ optical depth $\tau_\mathrm{eff}$ (measured over 50
    cMpc$/h$) in the simulation with measurements by
    \citet{2011MNRAS.410.1096B} and \citet{2018MNRAS.tmp.1287B}.
    Non-detections are shown as lower limits.  Upper limits in the
    measurements by \citet{2011MNRAS.410.1096B} indicate sightlines
    that do not have a formal detection but show individual
    transmission peaks.  The solid curve shows the mean
    $\tau_\mathrm{eff}$ evolution in the simulation, with the shaded
    region indicating the central 98\% extent.  The right panel shows
    in blue the 68\% scatter around the median cumulative distribution
    functions of $\tau_\mathrm{eff}$ of the simulated spectra at
    $z=5.0, 5.2, 5.4, 5.6, 5.8,$ and $6.2$ (left to right).  Also
    shown are the optimistic and pessimistic bounds measured by
    \citet{2018MNRAS.tmp.1287B} at $z\leq 5.8$.}
  \label{fig:tau_eff}
\end{figure*}

We assume that the ionizing luminosity of a source, $\dot N_\gamma$,
is proportional to its halo mass and require that the total volume
emissivity $\dot n=\sum\dot N_\gamma/V_\mathrm{box}$, where
$V_\mathrm{box}$ is the box volume, matches a pre-selected emissivity
evolution \citep[cf.][]{2015MNRAS.453.2943C}.  Our chosen emissivity,
shown in the left panel of Figure~\ref{fig:calibration}, peaks at
redshift $z\sim 6.8$ and drops towards higher redshifts somewhat more
slowly than current estimates of the evolution of the cosmic star
formation rate density.  This is strikingly different from the
evolution adopted by \citet{2015MNRAS.453.2943C} and
\citet{2018MNRAS.477.5501K}, in which the emissivity at $z>7$ is much
higher.  The emissivity assumed here drops by a factor of two between
$z=6.8$ and $z=4$.  As we discuss below, this decrease at $z<6.8$
allows us to reproduce the observed mean \lya\ transmission at these
redshifts, while the decrease in the emissivity towards higher
redshift at $z>6.8$ results in a rather late reionization.  Note that
the emissivity model required to match the \lya\ forest opacity as
well as the \citet{2018arXiv180706209P} Thompson scattering optical
depth is similar to the fiducial model for the ionizing emissivity of
galaxies in \citet{2018arXiv180104931P}.  As in the
\citet{2018arXiv180104931P} model, the difference in the evolutionary
trends of the emissivity and the cosmic star formation rate density
(Figure~\ref{fig:calibration}) can be attributed to the evolution of
the escape fraction of ionizing photons from galaxies, possibly due to
changes in morphology, stellar populations, and dust content of these
galaxies \citep{2018MNRAS.479..994R, 2017MNRAS.470..224T,
  2017MNRAS.466.4826K, 2015MNRAS.451.2544P, 2014ApJ...788..121K,
  2011MNRAS.412..411Y}.  We use a single photon frequency to reduce
the computational cost and assume that all sources have a blackbody
spectrum with $T=$70,000~K \citep{2018MNRAS.477.5501K}.  This yields
an average photon energy of 23.83~eV in the optically thick limit.  We
have varied these assumptions and found our results to be robust. We
will discuss more details in future work.

\section{Large \lya\ opacity fluctuations}
\label{sec:res}

Figure~\ref{fig:calibration} shows the evolution of the
volume-averaged ionized hydrogen fraction $Q_\mathrm{HII}$.
Reionization is considerably late in our simulation compared to most
models in the literature \citep[e.g.,][]{2012ApJ...746..125H}, and is
comparable to the `Very Late' model shown by
\citet{2015MNRAS.452..261C} to be consistent with the rapid
disappearence of \lya\ emission of high-redshift galaxies.  Half of
the cosmic volume is reionized at $z=7.0$.  The duration of
reionization, as quantified by the difference in the redshifts at
which 5\% and 95\% cosmic volume is reionized, $\Delta z \equiv
z_{5\%}-z_{95\%}$, is 3.89.  Reionization is complete at $z = 5.3$.
This evolution is also in excellent agreement with the determination
of $Q_\mathrm{HII}$ at $z=7.1$ and $7.5$ by
\citet{2018arXiv180206066D} and at $z=7.1$ by
\citet{2017MNRAS.466.4239G} from the red damping wing and the short
near-zones in the two highest redshift QSOs known.
Figure~\ref{fig:calibration} also shows the electron scattering
optical depth in our model, $\tau_{\rm CMB}=0.054$, in excellent
agreement with the most recent determination of $\tau_{\rm CMB}
=0.0544\pm 0.0073$ from \citet{2018arXiv180706209P}. Note that
following \citealt{2018arXiv180706209P}, we assume here that
\HeII\ reionizes instantaneously at $z=3.5$.

Figure~\ref{fig:slices} shows the neutral hydrogen fraction
$x_\mathrm{HI}$, gas temperature $T$, and the hydrogen photoionization
rate $\Gamma_\mathrm{HI}$ from $z=5$ to $8$.  These lightcones nicely
illustrate the patchy and delayed nature of reionization in our
simulation, with `islands' of neutral hydrogen several tens of
megaparsecs in length persisting down to $z<5.5$.  Large coherent
spatial variation of the neutral fraction are seen at even lower
redshifts ($z\sim 5$).  These are accompanied by large-scale, coherent
fluctuations (a factor of 3--4 at $z<5.5$) in the gas temperature that
persist all the way down to $z=4$ \citep[cf.][]{2018MNRAS.477.5501K}.
Note that the photoionisation rate is significantly reduced in the
vicinity of the remaining neutral islands likely due to a combination
of these regions only being recently ionized and having a reduced mean
free path for photons arriving from the direction of the neutral
islands.  Note further that the last neutral islands to be reionized
attain the highest temperatures and switch from exhibiting the largest
effective optical depth to exhibiting the lowest effective optical
depth at a given redshift very quickly.

Observations traditionally quantify effective optical depths over
spectral chunks corresponding to 50~cMpc$/h$.  At $z=5.8$, this
corresponds to $\Delta t\sim 35$~Myr.  As seen in
Figure~\ref{fig:slices}, the IGM evolves rapidly over this time scale
at these redshifts.  To incorporate this rapid evolution in our
simulated spectra, we interpolate lines of sight in time between
different snapshots.  Figure~\ref{fig:tau_eff} shows the resultant
evolution of \teff\ in our model in comparison with measurements from
\citet{2018MNRAS.tmp.1287B} and \citet{2015MNRAS.447.3402B}.  Our
simulated spectra match the data very well down to $z=4.9$.  The late
end of reionization and the persistence of large neutral hydrogen
islands down to $z<5.5$ result in sightlines that still have
$\tau_\mathrm{eff}>8$ at $z=5.7$.  Figure~\ref{fig:tau_eff} also shows
the cumulative distribution function of $\tau_\mathrm{eff}$ in six
redshift bins from $z=5$ to $5.8$ and at $z=6.2$.  At $z=5$--$5.8$, we
compare our results with the measurements by
\citet{2018MNRAS.tmp.1287B} who present their results as `optimistic'
and `pessimistic' limits on the distribution.  Lower limits on
\teff\ are treated as measurements in the optimistic case, whereas
these are assumed to have values greater than $\tau_\mathrm{eff}=8$ in
the pessimistic case.  When comparing the simulation with data, we
draw 50 samples with the same size and redshift distribution as that
of the data in \citet{2018MNRAS.tmp.1287B} and show the 68\% scatter
in Figure~\ref{fig:tau_eff}.  At $z=6.2$, where no measurements are
available yet, we used a sample size of 25.  The \teff\ values
reported by \citet{2018ApJ...864...53E} are systematically higher than
those measured by \citet{2018MNRAS.tmp.1287B}, but it is certainly
possible to match the \citet{2018ApJ...864...53E} data if we delay
reionization in our model further.  We also note that increasing the
spatial resolution does not change our results, although for a
factor-of-2 higher resolution we find that a small enhancement ($\sim
10$\%) in the emissivity is required as more high-density absorbers
are resolved.  Similarly, changing the photon energy in our simulation
does not affect the opacity fluctuations.  Since the simulation is
calibrated to the mean \lya\ transmission, a change in the photon
energy only marginally changes the ionization fraction and the
required ionizing photon emissivity without altering the opacity
distribution.

\section{Discussion and Conclusions}

We have presented a radiative transfer simulation of cosmic
reionization by galaxies that closely agrees with the measurements of
the effective Ly$\alpha$ opacity of the intergalactic medium at
$z=5$--$6$.  Our reionization history also agrees very well with the
electron scattering optical depth measurements from CMB experiments,
as well as constraints on the ionization state of the IGM from QSO
near-zones at $z\sim 7$ and the rapid disappearance of \lya\ emission
from high-redshift galaxies.  This very good agreement with a wide
range of data is owed to reionization occuring rather late in the
model with a rapid evolution in ionizing emissivity that is 
peaked at $z\sim 6.8$ suggesting that the contribution of galaxies to
the ionizing emissivity at $z\la 4$ is small.  The reionization
history in this model results in a broad scatter in the neutral
hydrogen fraction and large neutral hydrogen `islands' persisting to
redshifts as low as $z\lesssim 5.5$.  As we will discuss in future
work, these large late remaining neutral islands can result in long
Gunn-Peterson troughs resembling those seen in the data.  The
simulation also shows large spatially coherent fluctuations of the
temperature-density relation that persist down to $z<5$. Unlike other
proposed models, our modelling solves the mystery of the large scatter
of the \lya\ opacity on surprisingly large scales without requiring
ionizing sources or properties of the IGM in tension with current
observations and/or theoretical expectations.  Published \lya\ data
already now provides very tight constraints on the ionization and
thermal history of the IGM at $z\le 6$.  With further improved
\lya\ absorption/emission data and by adding information from
\lyb\ and metal absorption data it should soon be possible to fully
map out the exact history of the second half ($Q>0.5$) of cosmic
reionization or more.

\section*{Acknowledgements}

This work used the Cambridge Service for Data Driven Discovery (CSD3)
operated by the University of Cambridge (www.csd3.cam.ac.uk), provided
by Dell EMC and Intel using Tier-2 funding from the Engineering and
Physical Sciences Research Council (capital grant EP/P020259/1), and
DiRAC funding from the Science and Technology Facilities Council
(www.dirac.ac.uk).  This work further used the COSMA Data Centric
system operated Durham University on behalf of the STFC DiRAC HPC
Facility. This equipment was funded by a BIS National E-infrastructure
capital grant ST/K00042X/1, DiRAC Operations grant ST/K003267/1 and
Durham University. DiRAC is part of the National E-Infrastructure.  We
thank Joseph Hennawi and Anna-Christina Eilers for data and
discussions.  We acknowledge support from ERC Advanced Grant 320596
`Emergence'.

\bibliographystyle{mnras}
\bibliography{refs}

\bsp
\label{lastpage}
\end{document}